\documentclass[aps,prl,twocolumn,showpacs,byrevtex]{revtex4}

\usepackage{times,mathptm,gensymb}
\usepackage{graphicx,color}
\begin{document}

\title{Origin of Symmetric Dimer Images of Si(001) Observed in Low-Temperature STM}
\author{Xiao-Yan Ren$^{1,2}$, Hyun-Jung Kim$^{3,4}$, Chun-Yao Niu$^{1,5}$, Yu Jia$^{1,5{\dagger}}$, Jun-Hyung Cho$^{3,1,6*}$}
\affiliation{$^1$ International Laboratory for Quantum Functional Materials of Henan, and School of Physics and Engineering, Zhengzhou University, Zhengzhou 450001, China \\
$^2$ School of Mechanical and Electrical Engineering, Henan Institute of Science and Technology, Xinxiang 453003, China  \\
$^3$ Department of Physics and Research Institute for Natural Sciences, Hanyang University, 17 Haengdang-Dong, Seongdong-Ku, Seoul 133-791, Korea \\
$^4$ Korea Institute for Advanced Study, 85 Hoegiro, Dongdaemun-gu, Seoul 130-722, Korea \\
$^5$ Center for Advanced Analysis and Computational Science, Zhengzhou University, Zhengzhou 45001, China \\
$^6$ International Center for Quantum Design of Functional Materials (ICQD), HFNL, University of Science and Technology of China, Hefei, Anhui 230026, China}
\date{\today}

\begin{abstract}
It has been a long-standing puzzle why buckled dimers of the Si(001) surface appeared symmetric below ${\sim}$20 K in scanning tunneling microscopy (STM) experiments. Although such symmetric dimer images were concluded to be due to an artifact induced by STM measurements, its underlying mechanism is still veiled. Here, we demonstrate, based on a first-principles density-functional theory calculation, that the symmetric dimer images are originated from the flip-flop motion of buckled dimers, driven by quantum tunneling (QT). It is revealed that at low temperature the tunneling-induced surface charging with holes reduces the energy barrier for the flipping of buckled dimers, thereby giving rise to a sizable QT-driven frequency of the flip-flop motion. However, such a QT phenomenon becomes marginal in the tunneling-induced surface charging with electrons. Our findings provide an explanation for low-temperature STM data that exhibits apparent symmetric (buckled) dimer structure in the filled-state (empty-state) images.
\end{abstract}


\maketitle

Over the last 30 years the atomic and electronic structures of the Si(001) surface have been extensively investigated because of the fundamental building block for the fabrication of electronic devices as well as for the prototypical model system of semiconductor surfaces~\cite{Dabro,Chadi,Pollmann,yates,Medeiros,wol}. From enormous experimental and theoretical studies, it is well established that the basic reconstruction of Si(001) consists of the formation of buckled dimers~\cite{Tromp,Ram,Kipp,Shir}. However, at room temperature scanning tunneling microscopy (STM) experiments showed symmetric dimer images because of a thermally activated flip-flop motion of buckled dimers. Such apparent symmetric dimer images disappear below ${\sim}$120 K~\cite{Wolkow}, forming the $c(4{\times}2)$ or $p(2{\times}2)$ reconstruction structure [see Fig. 1(a)]. Surprisingly, further cooling below ${\sim}$20 K causes the buckled dimers to appear symmetric again~\cite{kondo,Yokoyama1}. Such symmetric-dimer STM images at low temperature have been explained in terms of various origins such as a dynamical flip-flop motion of buckled dimers~\cite{Yokoyama1,Mitsui}, local surface charging effects~\cite{Ono}, a possible asymmetric $p(2{\times}1)$ reconstruction~\cite{Perdigao}, and a contribution of bulk states~\cite{Hata,Manzano}. However, the microscopic mechanism underlying the low-temperature symmetric dimer images has remained an open question.

\begin{figure}[ht]
\centering{ \includegraphics[width=7.7cm]{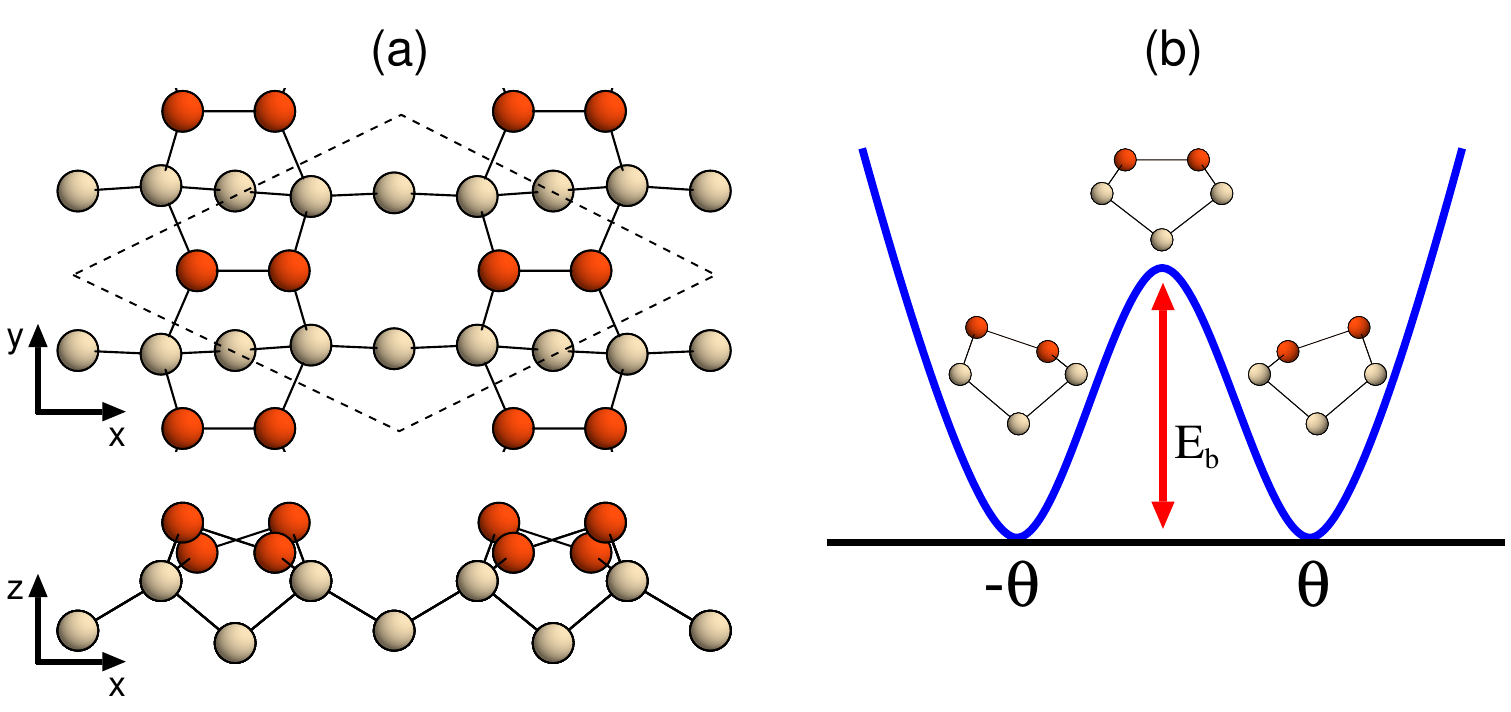} }
\caption{(Color on line) (a) Top and side views of the optimized $c$(4${\times}$2) structure of Si(001). The $c$(4${\times}$2) unit cell is indicated by the dashed line. The ${\bf x}$ (${\bf y}$) axis is perpendicular (parallel) to dimer rows, while the ${\bf z}$ axis is along the [001] direction. For distinction, the Si-dimer atoms are drawn with dark circles. In (b), the symmetric double-well potential for the flipping of buckled dimers is schematically drawn. Here, $E_b$ denotes the energy barrier, obtained by the energy difference between the $p$(2${\times}$1) and $c$(4${\times}$2) structures.}
\end{figure}

There have so far been a number of low-temperature STM experiments~\cite{Yokoyama1,Mitsui,Ono,Manzano} to characterize the apparent symmetric dimer images. Yokoyama and Takayanagi~\cite{Yokoyama1} observed that the symmetric dimer images measured at 5 K have flicker noise, which was explained by slow dynamical flip-flop motion of the buckled dimers during the STM scan. Mitsui and Takayanagi~\cite{Mitsui} found that at 65 K higher tunneling currents increase not only the area of symmetric dimer images but also the flip-flop rate of buckled dimers. Below 10 K, Ono $et$ $al$.~\cite{Ono} observed both buckled and symmetric dimer images depending on the polarity of the bias voltage: i.e., the buckled dimer images, locally forming $c$(4${\times}$2) or $p$(2${\times}$2) periodicity, were observed with positive bias voltages (empty-state images), while most of the dimers appear symmetric with negative bias voltages (filled-state images). Recently, Manzano $et$ $al$.~\cite{Manzano} reported that at 7 K the negative bias voltages smaller than $-$1.5 V remained a $c$(4${\times}$2) reconstruction, but those larger than $-$1.5 V produced symmetric dimer images. On the basis of existing low-temperature STM data~\cite{Yokoyama1,Mitsui,Ono,Manzano}, the following questions on the appearance of symmetric dimer images can be raised: i.e., Why does the activation barrier ($E_b$) for the flipping of buckled dimers become much reduced at low temperature? What is the reason why the filled-state and empty-state STM images exhibit symmetric and buckled dimer structures, respectively? How does the tunneling-induced surface charging at low temperature~\cite{Ono,Bri,Pol} or the electric field via bias voltage affect STM imaging to show apparent symmetric dimer structure?

In this Letter, we perform first-principles density-functional theory (DFT) calculations to investigate the energy difference (equivalently $E_b$) between the symmetric-dimer structure and the $c$(4${\times}$2) structure under electron or hole doping as well as in the presence of external electric field applied along the [001] direction. We find that, as the amount of hole doping increases, $E_b$ decreases more dominantly than the case of electron doping. Compared to such surface charging effects, the application of electric field is found to give a relatively small change in $E_b$. As $E_b$ decreases with hole doping, the thermally activated flipping rate of buckled dimers is still negligible below 10 K, but the quantum tunneling (QT) driven flip-flop motion can be enabled to produce the symmetric-dimer STM images. Such a QT phenomenon of buckled dimers is, however, marginal with electron doping. Thus, a long-standing puzzle about the appearance of symmetric dimer images in low-temperature STM experiments can be solved in terms of the QT-driven flip-flop motion of buckled dimers, which can be facilitated by the tunneling-induced surface charging with holes.

We begin to optimize both the symmetric dimer structure, forming a $p$(2${\times}$1) periodicity (hereafter, designated as the $p$(2${\times}$1) structure), and the $c$(4${\times}$2) structure by using the DFT calculation within the generalized-gradient approximation (GGA)~\cite{method}. The optimized $c$(4${\times}$2) structure is displayed in Fig. 1(a). We find that the $c$(4${\times}$2) structure consisting of alternatively buckled dimers along and perpendicular to the dimer rows has a dimer bond length of $d_D$ = 2.357 {\AA} and a dimer buckling angle of ${\theta}$ = 18.0$^{\circ}$. This $c$(4${\times}$2) structure is found to be more stable than the symmetric-dimer structure by 255 meV per dimer, yielding $E_b$ = 255 meV [see Fig. 1(b)]. As shown in Fig. 2(a) and 2(b), the calculated band structure of $p$(2${\times}$1) has a metallic band crossing the Fermi level $E_F$, whereas that of $c$(4${\times}$2) exhibits a semiconducting feature with a band gap of 0.27 eV. The present results for the geometry, energetics, and band structure of the $c$(4${\times}$2) structure are in good agreement with those of previous DFT calculations~\cite{Ram,Sei}.

\begin{figure}[ht]
\centering{ \includegraphics[width=7.7cm]{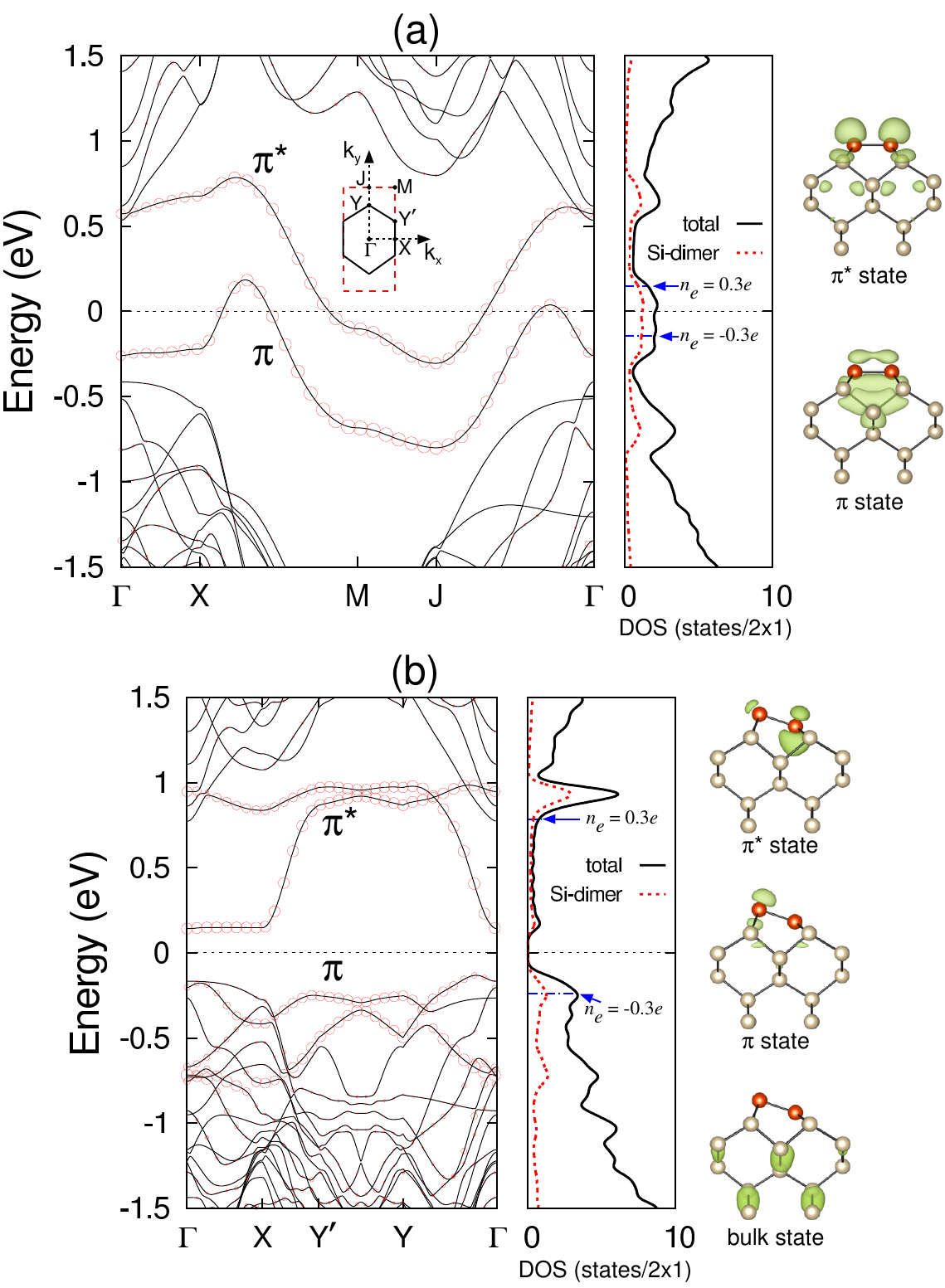} }
\caption{(Color on line) Calculated surface band structures of (a) the $p$(2${\times}$1) and (b) $c$(4${\times}$2) structures. The bands projected onto the $p_x$, $p_y$, and $p_z$ orbitals of Si-dimer atoms are displayed with circles whose radii are proportional to the weights of such orbitals. The energy zero represents $E_F$. The inset in (a)
shows the surface Brilloiun zones of the $p$(2${\times}$1) and $c$(4${\times}$2) unit cells. The total DOS and the local DOS of Si dimers are displayed with solid and dotted lines, respectively. The charge characters of the ${\pi}$ and ${\pi}^*$ surface states at the ${\Gamma}$ point are drawn with an isosurface of 0.05 $e$/{\AA}, while that of the bulk state of $c$(4${\times}$2) at the ${\Gamma}$ point (just below $E_F$) is drawn with an isosurface of 0.02 $e$/{\AA}.}
\end{figure}

It has been known that below ${\sim}$40 K electrons or holes, injected through tunneling current in STM, result in surface charging due to a slow carrier relaxation between the surface layer and the semiconducting bulk Si~\cite{Ono,Bri,Pol}. In order to examine the influence of surface charging on the energetics of the $p$(2${\times}$1) and $c$(4${\times}$2) structures, we perform total-energy calculations for the two structures with electron or hole doping. For the simulation of surface charging, we use the virtual crystal approximation~\cite{VCM} to compensate excess electrons $n_e$ or holes (whose amount is represented as a negative value of $n_e$). Figure 3 shows the calculated values of $E_b$ as a function of $n_e$ ranging from $-$0.6$e$ to 0.6$e$ per $p$(2${\times}$1) unit cell. We find that both the electron and hole dopings reduce the energy difference between the $p$(2${\times}$1) and $c$(4${\times}$2) structures. The resulting decrease of $E_b$ with electron or hole doping can be attributed to the metallic and semiconducting features of the $p$(2${\times}$1) and $c$(4${\times}$2) structures, respectively. As shown in Fig. 2(a) and 2(b), for instance of $|n_e|$ = 0.3$e$, excess electrons in $p$(2${\times}$1) are filled in the relatively lower unoccupied electronic states compared to those in $c$(4${\times}$2), whereas holes in $p$(2${\times}$1) are created in the relatively higher occupied electronic states compared to those in $c$(4${\times}$2). Therefore, as electron (hole) doping increases, the total energy of the $p$(2${\times}$1) [$c$(4${\times}$2)] structure is expected to decrease (increase) more largely compared to the $c$(4${\times}$2) [$p$(2${\times}$1)] structure, leading to a decrease of $E_b$.

As shown in Fig. 3, $E_b$ decreases more significantly with increasing hole doping, compared to the case of electron doping. This difference between electron and hole dopings may be due to the different characters of the unoccupied and occupied electronic states in the $c$(4${\times}$2) structure: i.e., the lowest unoccupied states are mostly the surface states of ${\pi}^*$ orbitals, while the occupied states below $E_F$ consist of the surface states of ${\pi}$ orbitals as well as the bulk states [see the total density of states (DOS) and the local DOS of Si dimers in Fig. 2(b)]. We note that, for hole doping with $n_e$ = $-$0.3$e$, the majority of the holes in the $c$(4${\times}$2) structure is created in the bulk states around the ${\Gamma}$ point (see Fig. 1S of the Supplemental Material~\cite{Supp}), possibly giving rise to a relatively larger strain energy compared to the $p$(2${\times}$1) structure where holes are created mostly in the surface states. This fact may cause a more significant decrease of $E_b$ with hole doping, compared to electron doping where both the $c$(4${\times}$2) and $p$(2${\times}$1) structures occupy excess electrons mostly in their surface states.

\begin{figure}[ht]
\centering{ \includegraphics[width=7.7cm]{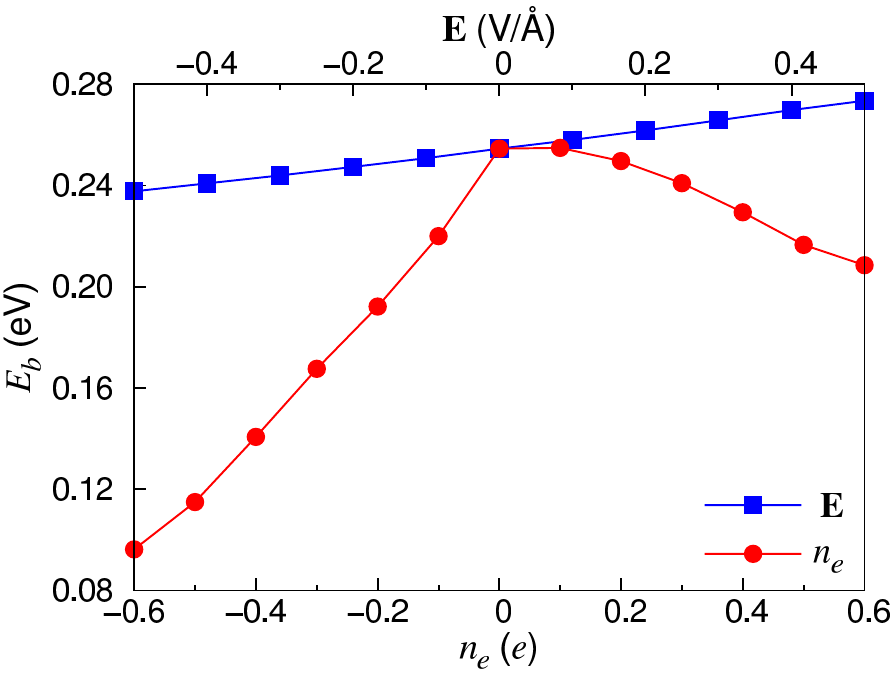} }
\caption{(Color on line) Calculated energy barrier $E_b$ [see Fig. 1(b)] for the flip-flop motion of buckled dimers as a function of electron and hole dopings as well as external electric field. The unit of $e$ in $n_e$ is given per $p$(2${\times}$1) unit cell.}
\end{figure}

Next, we examine the influence of external electric field ${\bf E}$ on the energetics of the $p$(2${\times}$1) and $c$(4${\times}$2) structures. Here, ${\bf E}$ is simulated by superimposing an additional sawtooth potential along the [001] direction (taken as the +$z$ direction) with discontinuity at the mid-plane of the vacuum region of the supercell. Note that an STM bias voltage of 1.5 V and a tip-sample distance of ${\sim}$5 {\AA} would give rise to an electrical field of ${\sim}$0.3 V/{\AA}. Figure 3 also shows the calculated values of $E_b$ as a function of ${\bf E}$ ranging between $-$0.5 and +0.5 V/{\AA}. We find that $E_b$ increases (decreases) as ${\bf E}$ increases along the +$z$ ($-z$) direction. These different behaviors of $E_b$ depending on the direction of ${\bf E}$ can be explained in terms of the different contributions of electrostatic energy due to external electric field between the $p$(2${\times}$1) and $c$(4${\times}$2) structures. Since the surface dipole moment p$_z$ (pointing $-z$ direction) of the metallic $p$(2${\times}$1) structure is larger in magnitude by 0.038 $e${\AA} than that of the semiconducting $c$(4${\times}$2) structure, an electric field applied along the +$z$ ($-z$) direction gives a positively (negatively) larger electrostatic energy $U$ = $-{\bf p}{\cdot}{\bf E}$ of surface dipole in $p$(2${\times}$1) compared to in $c$(4${\times}$2), leading to an increase (decrease) of $E_b$. We find that the variation of $E_b$ with respect to the external electric field of 0 ${\le}$ $|{\bf E}|$ ${\le}$ 0.5 V/{\AA} is less than ${\sim}$20 meV, much smaller than that (${\sim}$160 meV) obtained from hole doping (see Fig. 3). Thus, we can say that the influence of hole doping on $E_b$ is much more pronounced than that arising from external electric field.

To account for the symmetric dimer images observed from low-temperature STM experiments~\cite{Yokoyama1,Mitsui,Ono,Manzano}, we investigate the flip-flop motion of buckled dimers driven by either thermal activation~\cite{Hata2001} or quantum tunneling. For this, we assume a symmetric double-well potential [see Fig. 1(b)] that describes the potential energy surface of flipping dimers as a function of ${\theta}$. Using a harmonic approximation, we obtain a vibration frequency for this potential well as $f_0$ = ${\frac{\omega}{2{\pi}}}$ = ${\frac{1}{2{\pi}}}\sqrt{\frac{k}{I}}$ ${\approx}$ 0.3${\times}10^{13}$ sec$^{-1}$ in the absence of electron or hole doping, where the torsion constant $k$ and the inertia moment $I$ of flipping dimer can be estimated from $E_b$ = $k{{\theta}_0}^2$ (${\theta}_0$: dimer buckling angle at the lowest-energy configuration) and $I$ = $\frac{1}{2}m_{\rm si}{d_D}^2$ ($m_{\rm si}$: mass of Si atom). Based on an Arrhenius-type activation process, a thermally excited flipping rate can be expressed as $f_{T}$ = $f_{0}exp(\frac{-E_b}{kT})$. With the calculated values of $E_b$ and $f_0$ as a function of $|n_e|$ ${\le}$ 0.6$e$, we obtain $f_T$ smaller than 0.8${\times}10^{-36}$ sec$^{-1}$ at 10 K. This thermal flipping rate is too small to explain the observed symmetric-dimer STM images with flicker noise~\cite{Yokoyama1,Mitsui}. As an alternative explanation for the flip-flop motion of buckled dimers, we consider quantum tunneling (QT) within the double-well potential, whose flipping rate can be approximated~\cite{Fab,Mer} as
\begin{eqnarray}
f_{\rm QT} = \sqrt{\frac{8E_b{\omega}}{{\hbar}{\pi}}} exp({\frac{-2E_b}{{\hbar}{\omega}}}).
\end{eqnarray}
Contrasting with $f_{T}$, $f_{\rm QT}$ is independent of temperature, while it is determined by the ratio of $E_b$ and the zero-point energy ${\frac{1}{2}}{\hbar}{\omega}$. The estimated values of $f_{\rm QT}$ are plotted as a function of $n_e$. We find that $f_{\rm QT}$ sharply increases with increasing hole doping, while it is nearly flat with respect to electron doping. Here, note that electron doping decreases ${\omega}$ due to an increase of ${\theta}_0$, thereby hardly changing the ratio of $E_b$ and ${\frac{1}{2}}{\hbar}{\omega}$. For hole doping with $|n_{e}|$ $>$ 0.3$e$, $f_{\rm QT}$ becomes greater than ${\sim}$1.9${\times}10^{-1}$ sec$^{-1}$. Considering that it takes about 10$^{-2}$ sec to obtain an STM image of a dimer, such a hole-doping induced flip-flop motion can produce the observed symmetric dimer images in low-temperature STM experiments~\cite{Yokoyama1,Mitsui,Ono,Manzano}. It is noticeable that the application of ${\bf E}$ along the $-z$ (+$z$) direction decreases (increases) $E_b$. Consequently, one expects that negative sample bias (equivalently, negative electric field) inducing hole doping at low temperature enhances the magnitude of $f_{\rm QT}$. On the other hand, positive sample bias (positive electric field) inducing electron doping suppresses $f_{\rm QT}$. These drastically different aspects of negative and positive bias voltages in low-temperature STM experiments account for the observations of symmetric and buckled dimer images in filled-state and empty-state images, respectively~\cite{Ono,Manzano}.

\begin{figure}[ht]
\centering{ \includegraphics[width=7.7cm]{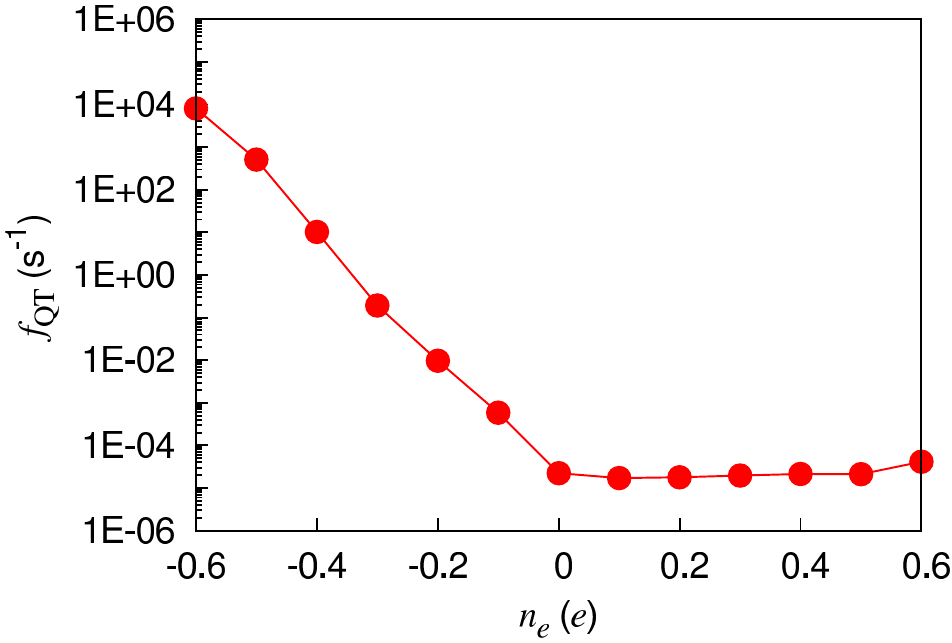} }
\caption{(Color on line) Calculated QT-driven flipping rate of buckled dimers as a function of electron and hole dopings.}
\end{figure}

Although we present a simple picture of the QT-driven flip-flop motion of buckled dimers with a double-well potential, we believe that it captures the microscopic mechanism underlying low-temperature symmetric-dimer STM images, as explained above. It is noted that the present DFT-GGA calculation may tend to somewhat overestimate the energy gain due to buckling. Indeed, the quantum Monte Carlo (QMC) calculation~\cite{QMC} which accurately describes electronic correlations extrapolates the value of $E_b$ up to ${\sim}$150 meV per dimer. Obviously, this reduction of $E_b$ should enhance the QT-driven flip-flop motion of buckled dimers. More rigorous QMC simulations with sufficiently large clusters or slab geometries will be a subject of future work.

In summary, we have performed a DFT-GGA calculation for the Si(001) surface to investigate the energy difference between the symmetric-dimer structure and the $c$(4${\times}$2) structure under electron or hole doping as well as applied external electric field along the [001] direction. This energy difference corresponding to the energy barrier for the flipping of buckled dimers was found to decrease more significantly with respect to hole doping compared to electron doping. Consequently, we found that hole doping gives rise to a sizable QT-driven frequency of the flip-flop motion of buckled dimers while electron doping shows the marginal QT effects. These different QT aspects of hole and electron dopings are most likely to yield the imaging difference between the filled- and empty-state STM images at low temperature. Thus, we concluded that quantum tunneling enhanced by the tunneling-induced hole doping causes the observation of symmetric dimer images in low-temperature STM experiments.

\noindent {\bf ACKNOWLEDGEMENTS}
We thank Prof. Changgan Zeng, Prof. Shun-Fang Li, and Prof. Zhenyu Zhang for helpful discussions. This work was supported by the National Basic Research Program of China (Grant No. 2012CB921300), National Natural Science Foundation of China (Grants No. 11274280 and No. 61434002), and National Research Foundation of Korea (Grant No. 2015R1A2A2A01003248). The calculations were performed by KISTI supercomputing center through the strategic support program (KSC-2015-C3-017) for the supercomputing application research.

\noindent $^{*}$Corresponding author: chojh@hanyang.ac.kr \\
\noindent $^{\dagger}$Corresponding author: jiayu@zzu.edu.cn


\end{document}